\documentclass[%
 aip,
 amsmath,amssymb,
 reprint,%
]{revtex4-1}

\usepackage{graphicx}
\usepackage{dcolumn}
\usepackage{bm}
\usepackage{bm} 
\usepackage[colorlinks=true,
            linkcolor=blue,
            urlcolor=blue,
            citecolor=blue]{hyperref}
\usepackage[utf8]{inputenc}
\usepackage[T1]{fontenc}
\usepackage{mathptmx}
\usepackage{etoolbox}
\usepackage{physics}

\makeatletter
\def\@email#1#2{%
 \endgroup
 \patchcmd{\titleblock@produce}
  {\frontmatter@RRAPformat}
  {\frontmatter@RRAPformat{\produce@RRAP{*#1\href{mailto:#2}{#2}}}\frontmatter@RRAPformat}
  {}{}
}%
\makeatother

\makeatletter
\newcommand{\vast}{\bBigg@{4}}
\newcommand{\Vast}{\bBigg@{5}}
\makeatother

\begin{document}

    \preprint{AIP/123-QED}

    \title{An investigation of escape and scaling properties of a billiard system}
    \author{Matheus Rolim Sales}
    \email{matheusrolim95@gmail.com}
    \affiliation{Departamento de Física, Universidade Estadual Paulista (UNESP), 13506-900, Rio Claro, SP, Brasil}
    \author{Daniel Borin}
    \affiliation{Departamento de Física, Universidade Estadual Paulista (UNESP), 13506-900, Rio Claro, SP, Brasil}
    \author{Diogo Ricardo da Costa}
    \affiliation{Departamento de Física, Universidade Estadual Paulista (UNESP), 13506-900, Rio Claro, SP, Brasil}
    \author{José Danilo Szezech Jr.}
    \affiliation{Programa de Pós-Graduação em Ciências, Universidade Estadual de Ponta  Grossa, 84030-900, Ponta Grossa, PR, Brasil}
    \affiliation{Departamento de Matemática e Estatística, Universidade Estadual de Ponta Grossa, 84030-900, Ponta Grossa, PR, Brasil}
    \author{Edson Denis Leonel}
    \affiliation{Departamento de Física, Universidade Estadual Paulista (UNESP), 13506-900, Rio Claro, SP, Brasil}
    \date{\today}

    \begin{abstract}
        We investigate some statistical properties of escaping particles in a billiard system whose boundary is described by two control parameters with a hole on its boundary. Initially, we analyze the survival probability for different hole positions and sizes. We notice the survival probability follows an exponential decay with a characteristic power law tail when the hole is positioned partially or entirely over large stability islands in phase space. We find the survival probability exhibits scaling invariance with respect to the hole size. In contrast, the survival probability for holes placed in predominantly chaotic regions deviates from the exponential decay. We introduce two holes simultaneously and investigate the complexity of the escape basins for different hole sizes and control parameters by means of the basin entropy and the basin boundary entropy. We find a non-trivial relation between these entropies and the system's parameters and show that the basin entropy exhibits scaling invariance for a specific control parameter interval.
    \end{abstract}
    \keywords{Classical billiards, escape of particles, survival probability, scaling invariance, basin entropy}

    \maketitle

    \begin{quotation}
        
        The phase space of a typical two-dimensional Hamiltonian system is not completely ergordic. There is a coexistence of chaotic and regular regions that gives rise to the well-known phenomenon of stickiness. Chaotic orbits become trapped near stability islands for long, but finite, times, and this intermitence in the chaotic motion shapes the transport and statistical properties across phase space. In this paper, we analyze the escape dynamics of a billiard system whose boundary is defined by two control parameters with an exit hole along its boundary. We find the survival probability either follows an exponential or a stretched exponential decay depending on the position of the hole. By introducing two holes simultaneously, we construct the escape basins for different hole's sizes and quantify the basins complexity using the basin entropy and the basin boundary entropy. The complexity of the basins depends nontrivially on the control parameters and we find that the basin entropy exhibits scaling invariance for a specific control parameter interval.


    \end{quotation}

    \section{Introduction}

    In general, the phase space of a typical quasi-integrable Hamiltonian system is mixed, where regular and chaotic domains coexist~\cite{lichtenberg2013regular}. The regular regions consist of periodic and quasiperiodic orbits that lie on invariant tori, while the chaotic orbits fill densely the whole available region in phase space. For two-dimensional area-preserving maps, the invariant tori divides the phase space into distinct and unconnected domains, \textit{i.e.}, an orbit initially inside of an island will never reach the chaotic sea and vice versa~\cite{lichtenberg2013regular,Mackay1984}. The stability islands and chaotic regions organize themselves in phase space in an infinite hierarchical islands-around-islands structure, where the larger islands are surrounded by smaller islands, which are in turn surrounded by even smaller islands and so on for increasingly smaller scales~\cite{Meiss1985, meiss_transport}. This complex interplay between stability islands and chaotic regions gives rise to the phenomenon of stickiness~\cite{Contopoulos1971, KARNEY1983360, MEISS1983375, CHIRIKOV1984395, Efthymiopoulos_1997, Contopoulos2008, Cristadoro2008, Contopoulos2010}. The stickiness of chaotic orbits occurs near stability islands and these orbits experience long, but finite, periods of nearly quasiperiodic motion. Before escaping to the chaotic sea, these orbits are trapped within regions bounded by cantori~\cite{Mackay1984,Mackay1984b,meiss_transport,Efthymiopoulos_1997}. The cantori, which are a Cantor set, formed by the remnants of the destroyed Kolmogorov-Arnold-Moser (KAM) tori, as predicted by the KAM theorem~\cite{lichtenberg2013regular}, have a different function in the transport of particles in phase space than the KAM tori. While the KAM tori divide the phase space into distinct regions, the cantori act as partial barriers to the transport in phase space. The orbits may be trapped in a region bounded by the cantori, and once inside a cantorus, the chaotic orbits may transition to an inner cantorus, and so on, to arbitrarily small levels in the hierarchical structure of islands-around-islands.

    The stickiness affects the statistical properties of the system, such as the decay of correlations~\cite{KARNEY1983360,MEISS1983375,Vivaldi1983, CHIRIKOV1984395, Lozej2018} and transport~\cite{Zaslavski1972, Zaslavsky1997, ZASLAVSKY2002461, zaslavsky2005hamiltonian}. For closed systems, the transport properties may be studied considering the recurrence-time statistics (RTS)~\cite{Afraimovich1997, Altmann2005,Tanaka2006, Altmann2006,Venegeroles2009, Abud2013, lozej2020}, while for open systems it is customary to analyze the survival probability~\cite{Lai1992, Cristadoro2008,Altmann2009, Avetisov2010, DETTMANN2012403, LEONEL20121669,Livorati2012, Livorati2014,Mendez-Bermudez_2015,deFaria2016, LIVORATI2018225,BORIN2023113965}. For both cases, strongly chaotic dynamics leads to an asymptotic exponential decay, while in systems that exhibit stickiness, a power law tail emerges. Whether the decay follows an exponential or power law corresponds to normal or anomalous transport~\cite{Zaslavski1972, Zaslavsky1997, ZASLAVSKY2002461, zaslavsky2005hamiltonian}, respectively.

    In this paper, we study the escaping properties of a billiard system with static boundary and the scaling invariance~\cite{scalinglaws} of some observables. Essentially, when a system exhibits scaling invariance, its expected behavior remains consistent and robust regardless of scale. It is explored in various systems ranging from area-preserving maps, dissipative maps and billiards as well~\cite{Scaling1, Scaling2, Scaling3, Scaling4,Scaling5, Scaling6, Scaling7, Scaling8, Scaling9}, and more recently it has been explored for fractional maps~\cite{mendezbermudez2024scaling,BORIN2024114597}. The billiard system with static boundary is a Hamiltonian system and it is one of the simplest dynamical systems to exhibit chaotic motion. In its two-dimensional formulation, a point-like particle is confined to a planar region $\Omega$ delimited by hard walls $\partial\Omega$. The particle undergoes elastic collisions with the boundary $\partial\Omega$ such that the angle of incidence equals the angle of reflection~\cite{tabachnikov2005geometry}. Because billiards have a relatively simple structure, whether chaotic behavior emerges is entirely determined by the geometric characteristics of $\partial\Omega$, \textit{i.e.}, the presence of dispersing or defocusing components in the boundary $\partial\Omega$~\cite{Bunimovich_2018}. Therefore, different billiard geometries yield different dynamical behavior, namely, fully regular~\cite{tabachnikov2005geometry}, in which all orbits lie on periodic or quasiperiodic tori, fully chaotic~\cite{Sinai_1970,Bunimovich1974,Robnik_1983}, in which almost every orbit fills densely the entire phase space, and, mixed dynamics~\cite{Altmann2005,Altmann2006,Bunimovich2001, DACOSTA2020105440}, where the phase space is composed of both regular and chaotic domains, typical of quasi-integrable Hamiltonian systems. Billiard systems have also been studied in the context of quantum~\cite{Berry_1989, CASATI1999293, Barnett2007, deMenezes2007, ZANETTI20081644} and relativistic~\cite{Deryabin2003, Deryabin2004,PINTO20113273} mechanics.

    We consider in this paper a billiard system whose boundary depends on two control parameters. This system has been introduced in the context of quantum mechanics~\cite{billiard1} and recently some of its classical dynamical properties have been studied~\cite{billiard2}. Our focus lies in examining the escaping properties of an ensemble of particles through a hole placed on the billiard boundary. We analyze the survival probability for different hole positions and hole sizes as well. We find that when the hole overlaps, either partially or entirely, with larger stability islands, the survival probability follows an exponential decay with a characteristic power law tail. Also, in these cases, the survival probability exhibits scaling invariance with respect to the hole size. On the other hand, when the hole is placed within a predominantly chaotic region of phase space, the survival probability deviates from this exponential decay. We extend our analysis by introducing two holes simultaneously, and we construct the escape basins for different hole sizes. We find the escape basins to be more complex, in the sense of having fewer definite structures, for smaller hole sizes. We quantify this complexity by means of the basin entropy, $S_b$, and the basin boundary entropy, $S_{bb}$~\cite{Daza2016,Daza2017}. We find that the larger the hole size, the smaller both entropies become, doing so in a non-trivial and intricate manner. Nonetheless, we find that for a specific parameter interval, $S_b$ has an exponential dependence on the control parameter. Additionally, we show that $S_b$ also exhibits scaling invariance relative to this control parameter.

    This paper is organized as follows. In Section~\ref{sec:model} we formally introduce billiard systems and the system under study in this paper. We also demonstrate the algorithm used to calculate the successive collisions of the particle with the billiard boundary. In Section~\ref{sec:survivalprobability} we introduce one hole on the billiard boundary from where the particles can escape. We calculate the survival probability for several hole sizes and positions and show that the survival probability exhibits scaling invariance when the hole is placed partially or entirely over the large stability islands. In Section~\ref{sec:escapebasins} we consider two holes open simultaneously and construct escape basins for different hole sizes. We characterize the basins by means of the basin entropy and show that the basin entropy depends non-trivially on the hole sizes and the billiard parameters. We also show that the basin entropy exhibits scaling invariance. Section~\ref{sec:finalremarks} contains our final remarks.

    \section{Model and mapping}
    \label{sec:model}

    In the two-dimensional formulation of billiards, one considers a point-like particle of mass $\mu$, or an ensemble of particles, confined in a simply connected planar region $\Omega$ delimited by hard walls $\partial\Omega$. A billiard system with static boundary is a Hamiltonian system with potential $V\qty(\vb{q}) \equiv 0$ within the boundary and infinity on the boundary $\partial\Omega$, \textit{i.e.}, its Hamiltonian function is given by
    \begin{equation}
        \mathcal{H}(\vb{p}, \vb{q}) = \frac{\vb{p}^2}{2\mu} + V(\vb{q}),
    \end{equation}
    with
    \begin{equation}
        V(\vb{q}) = \Bigg\{\begin{array}{@{}l@{}l}
            0,& \text{for }\vb{q}\in\Omega,\\
            \infty,& \text{ otherwise},
        \end{array}
    \end{equation}
    where $\vb{p}$ and $\vb{q}$ are the generalized momentum and position, respectively. The particles undergo elastic collisions with the boundary such that only the momentum's direction is changed and the total mechanical energy of the system, $\mathcal{H}(\vb{p}, \vb{q}) \equiv E = \vb{p}^2/2\mu + V(\vb{q})$, is a constant of motion. Also, the angle of incidence equals the angle of reflection~\cite{tabachnikov2005geometry}.

    \begin{figure}[tb]
        \centering
        \includegraphics[width=\linewidth]{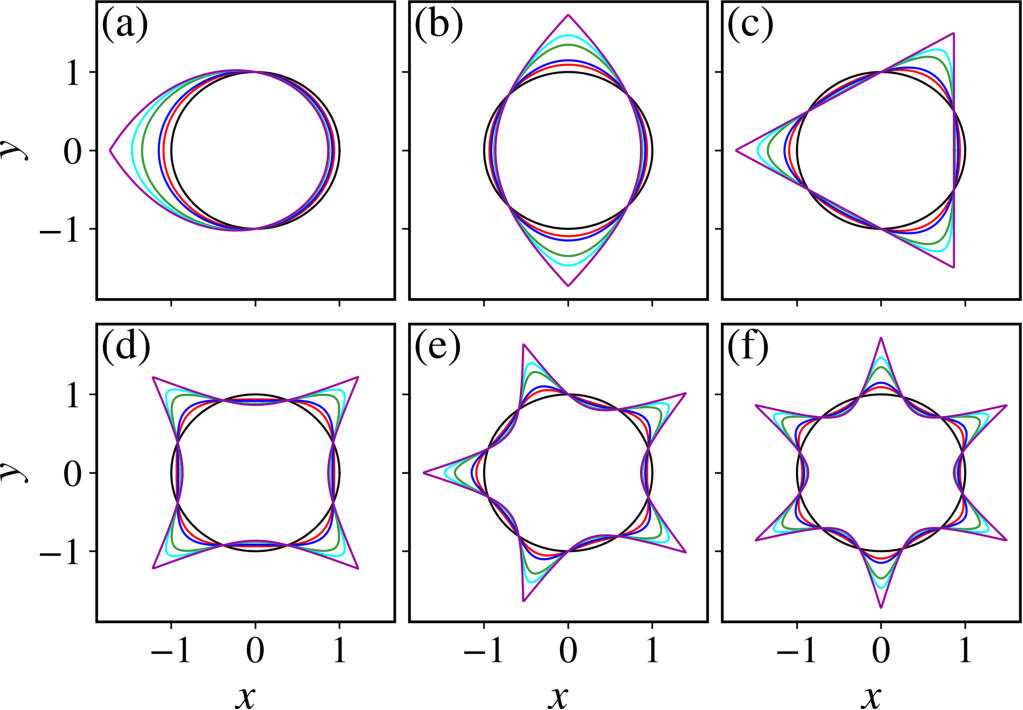}
        \caption{The billiard boundary for (a) $\gamma = 1$, (b) $\gamma = 2$, (c) $\gamma = 3$, (d) $\gamma = 4$, (e) $\gamma = 5$, and (f) $\gamma = 6$ with different values of $\xi$, namely, (black) $\xi = 0.0$, (red) $\xi = 0.15$, (blue) $\xi = 0.30$, (green) $\xi = 0.75$, (cyan) $\xi = 0.90$, and (purple) $\xi = 0.99999$.}
        \label{fig:shapes}
    \end{figure}

    \begin{figure*}[tb]
        \centering
        \includegraphics[width=\linewidth]{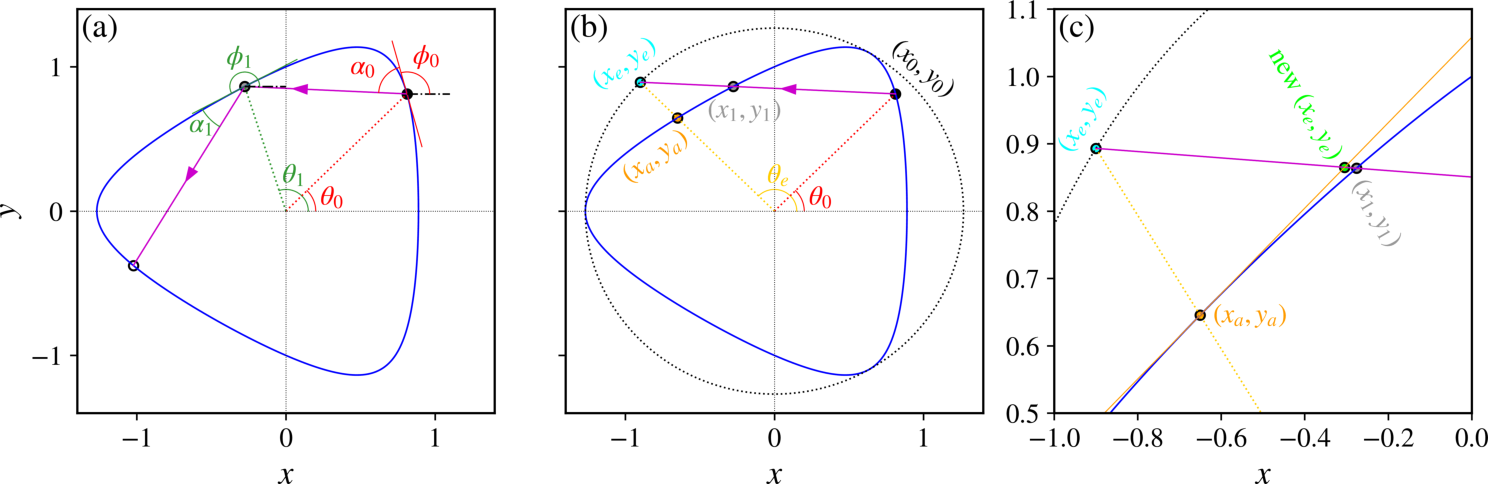}
        \caption{(a) Illustration of the billiard boundary and the angles considered in the billiard map for $\gamma = 3$ and $\xi = 0.6$ and initial condition $\qty(\theta_0, \alpha_0) = \qty(\pi/4, 2\pi/5)$. Panels (b) and (c) illustrate the algorithm for finding the next collision point, as discussed in the main text.}
        \label{fig:angles}
    \end{figure*}

    In this paper, we study a family of billiards with the boundary radius, $R\qty(\theta)$, implicitly parameterized by~\cite{billiard1,billiard2}
    \begin{equation}
        \label{eq:radius}
        R^2 + \frac{2\sqrt{3\xi}}{9}R^3\cos\qty(\gamma\theta) = 1,
    \end{equation}
    where $\theta \in [0, 2\pi)$ is the polar angle measured counterclockwise from the horizontal axis, $\gamma$ is an integer number, and $\xi \in [0, 1)$ controls the shape of the boundary. Figure~\ref{fig:shapes} displays different boundary shapes for different parameter values. $\xi = 0$ yields a circular shape for all $\gamma$ and $\gamma = 3$ and $\gamma = 4$ yields an equilateral triangle and a square-like shape, respectively, for $\xi \rightarrow 1$. The case $\gamma = 3$ is particularly interesting because both $\xi = 0$ and $\xi \rightarrow 1$ yield fully integrable billiard shapes.

    The billiard map is a two-dimensional nonlinear mapping $\mathbb{M}:\mathbb{R}^2 \rightarrow \mathbb{R}^2$. We characterize the particle's collisions with the boundary by two angles: $\theta$ and $\alpha$. The mapping relates these variables before and after the $n$th-collision
    \begin{equation}
        \qty(\theta_{n + 1}, \alpha_{n + 1}) = \mathbb{M}\qty(\theta_n, \alpha_n) = \mathbb{M}^n\qty(\theta_0, \alpha_0),
    \end{equation}
    where $\theta$ is the polar angle and $\alpha \in [0, \pi]$ is measured counterclockwise from the tangent line at the collision point and it is a complementary angle that measures the particle's direction of motion from the tangent line. Considering a particle initially at $\theta_n$ with initial angle $\alpha_n$, the particle starts its motion from the initial point $\qty(x_n, y_n)$ given by, in Cartesian coordinates,
    \begin{equation}
        \begin{aligned}
            x(\theta_n) \equiv x_n &= R\qty(\theta_n)\cos\theta_n,\\
            y(\theta_n) \equiv y_n &= R\qty(\theta_n)\sin\theta_n.\\
         \end{aligned}
    \end{equation}
    It is convenient to define the slope $\phi$ of the tangent line measured counterclockwise from the horizontal axis as well. It is given by
    \begin{equation}
        \phi_n = \arctan\qty[\frac{y'\qty(\theta_n)}{x'\qty(\theta_n)}]\mod{2\pi},
    \end{equation}
    where the prime indicates the derivative with respect to $\theta$. Therefore, the direction of the particle's momentum, measured counterclockwise from the horizontal axis, is
    \begin{equation}
        \mu_n = \alpha_n + \phi_n\mod{2\pi}.
    \end{equation}

    Since no forces are acting on the particle between two subsequent collisions the particle follows a straight line described by the following equations:
    \begin{equation}
        \label{eq:collisions}
        \begin{aligned}
            x_{n + 1} &= x_n + v_n\cos\qty(\mu_n)\Delta{t},\\
            y_{n + 1} &= y_n + v_n\sin\qty(\mu_n)\Delta{t},
        \end{aligned}
    \end{equation}
    where $\Delta{t}$ is the time interval between two collisions. We consider $v_n = 1$ without loss of generality and the particle's trajectory is given by
    \begin{equation}
        y\qty(\theta_{n + 1}) - y\qty(\theta_n) = \tan\qty(\mu_n)\qty[x\qty(\theta_{n + 1}) - x\qty(\theta_n)],
    \end{equation}
    where $\theta_{n + 1}$ is the new angular position of the particle where it hits the boundary. The direction of the particle's trajectory after the collision is given by
    \begin{equation}
        \alpha_{n + 1} = \phi_{n + 1} - \mu_n\mod{\pi}.
    \end{equation}
    Therefore, the final form of the mapping $\mathbb{M}$ is
    \begin{equation}
        \label{eq:mapping}
        \mathbb{M}:\vast\{\begin{array}{@{}l@{}l}
            F\qty(\theta_{n + 1}) &= y\qty(\theta_{n + 1}) - y\qty(\theta_n) - \\
            &-\tan\qty(\mu_n)\qty[x\qty(\theta_{n + 1}) - x\qty(\theta_n)] = 0,\\
            \alpha_{n + 1} &= \phi_{n + 1} - \mu_n\mod{2\pi}.
        \end{array}
    \end{equation}
    
    Figure~\ref{fig:angles}(a) shows the angles mentioned above for two subsequent collisions. Usually, the angle $\theta_{n + 1}$ is obtained numerically from $F\qty(\theta_{n  + 1}) = 0$ using a bisection method~\cite{LEONEL20121669}, for example. However, in our case, we consider a more efficient algorithm to calculate $\theta_{n + 1}$~\cite{billiard2,tangentmethod}, which we outline shortly. This algorithm can be 25 times faster in some situations than the traditional algorithm for studying billiards~\cite{tangentmethod}, and it is illustrated in Figures~\ref{fig:angles}(b) and~\ref{fig:angles}(c). It is important to note that even though this is an efficient algorithm, it is not applicable when the boundary has convex components. In our model, the billiard shapes have no convex components for $\gamma \leq 3$ (see Figure~\ref{fig:shapes}), and we limit our analysis to $\gamma = 3$. For an extended and more general version of this algorithm, we refer the reader to Ref.~\cite{tangentmethod}.

        \begin{figure*}[tb]
            \centering
            \includegraphics[width=\linewidth]{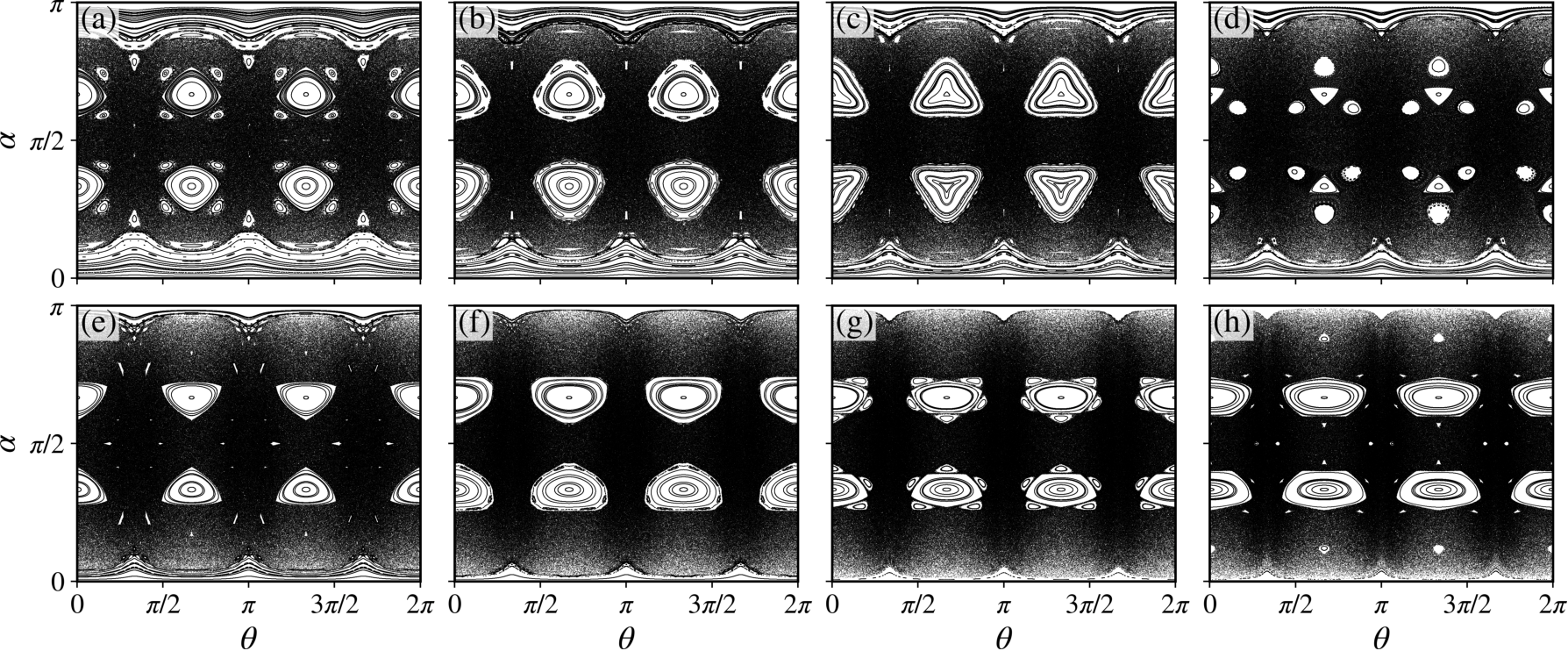}
            \caption{The phase space for $\gamma = 3$ and (a) $\xi = 0.20$, (b) $\xi = 0.30$, (c) $\xi = 0.40$, (d) $\xi = 0.45$, (e) $\xi = 0.55$, (f) $\xi = 0.70$, (g) $\xi = 0.80$, and (h) $\xi = 0.85$.}
            \label{fig:phase_space}
        \end{figure*}

    First, we consider an external circle to the billiard boundary with radius $R_{\mathrm{max}} = R\qty(\pi/\gamma)$ [dotted black line in Figures~\ref{fig:angles}(b) and~\ref{fig:angles}(c)]. The time it takes for the particle, initially at $\qty(x_0, y_0)$ [black dot in Figure~\ref{fig:angles}(b)], to reach the outer circle is obtained from $x_p^2 + y_p^2 = R_{\mathrm{max}}^2$ [cyan dot in Figures~\ref{fig:angles}(b) and~\ref{fig:angles}(c)], where $x_p$ and $y_p$ are given by Eqs.~\eqref{eq:collisions}. Thus, we obtain a quadratic equation for $\Delta{t}$
    \begin{equation}
        \begin{aligned}
            \qty(\Delta{t})^2 + 2\qty[x_0\cos\mu_0 + y_0\sin\mu_0]\Delta{t} + \\
            + x_0^2 + y_0^2 - R_{\mathrm{max}}^2 = 0,
        \end{aligned}
    \end{equation}
    with solution
    \begin{equation}
        \Delta{t}_e = \frac{-b + \sqrt{b^2 - 4c}}{2},
    \end{equation}
    where
    \begin{equation}
        \begin{aligned}
            b &= 2\qty[x_0\cos\mu_0 + y_0\sin\mu_0], \\
            c &= x_0^2 + y_0^2 - R_{\mathrm{max}}^2.
        \end{aligned}
    \end{equation}
    Hence, the Cartesian coordinates $(x_e, y_e)$ of the particle's collision point with the outer circle [cyan dot in Figures~\ref{fig:angles}(b) and~\ref{fig:angles}(c)] and its angular position are, respectively,
    \begin{equation}
        \begin{aligned}
            x_e &= x_0 + \cos\qty(\mu_0)\Delta{t_e},\\
            y_e &= y_0 + \sin\qty(\mu_0)\Delta{t_e},\\
            \theta_e &= \arctan\qty(\frac{y_e}{x_e})\mod{2\pi}.
        \end{aligned}
    \end{equation}
    We proceed to find the position on the billiard boundary for the angle $\theta_e$ [orange dot in Figures~\ref{fig:angles}(b) and~\ref{fig:angles}(c)]
    \begin{equation}
        \begin{aligned}
            x_a &= R\qty(\theta_e)\cos\theta_e,\\
            y_a &= R\qty(\theta_e)\sin\theta_e,
        \end{aligned}
    \end{equation}
    and the tangent line that passes through this point $(x_a, y_a)$ [orange line in Figure~\ref{fig:angles}(c)]
    \begin{equation}
        y_t(x) = y_a + \frac{y'\qty(\theta_e)}{x'\qty(\theta_e)}\qty(x - x_a).
    \end{equation}
    Next, we calculate the interception of this tangent line with the particle's trajectory [lime green dot in Figure~\ref{fig:angles}(c)] as $y_p = y_t(x_p)$:
    \begin{equation}
        \begin{aligned}
            &y_0 + \sin\qty(\mu_0)\Delta{t_e^{\mathrm{new}}} = \\
            &= y_a + \frac{y'\qty(\theta_e)}{x'\qty(\theta_e)}\qty[x_0 + \cos\qty(\mu_0)\Delta{t_e^{\mathrm{new}}} - x_a].
        \end{aligned}
    \end{equation}
    Isolating $\Delta{t_e^{\mathrm{new}}}$, we obtain
    \begin{equation}
        \Delta{t_e^{\mathrm{new}}} = \frac{y_a - y_0 + \frac{y'\qty(\theta_e)}{x'\qty(\theta_e)}\qty(x_0 - x_a)}{\sin\qty(\mu_0) - \frac{y'\qty(\theta_e)}{x'\qty(\theta_e)}\cos\qty(\mu_0)}.
    \end{equation}
    Therefore, the new interception point $\qty(x_e^{\mathrm{new}}, y_e^{\mathrm{new}})$ [lime green dot in Figure~\ref{fig:angles}(c)] and its angular position are given by, respectively,
    \begin{equation}
        \begin{aligned}
            x_e^{\mathrm{new}} &= x_0 + \cos\qty(\mu_0)\Delta{t_e^{\mathrm{new}}},\\
            y_e^{\mathrm{new}} &= y_0 + \sin\qty(\mu_0)\Delta{t_e^{\mathrm{new}}},\\
            \theta_e^{\mathrm{new}} &= \arctan\qty(\frac{y_e^{\mathrm{new}}}{x_e^{\mathrm{new}}}).
        \end{aligned}
    \end{equation}
    
    If $\abs*{\theta_e^{\mathrm{new}} - \theta_e} < \mathrm{TOL}$, $\abs*{x_e^{\mathrm{new}} - x_a} < \mathrm{TOL}$, and $\abs*{y_e^{\mathrm{new}} - y_a} < \mathrm{TOL}$, with $\mathrm{TOL} = 10^{-11}$, we consider $\theta_e^{\mathrm{new}}$ as the angular position of the particle's collision with the billiard boundary, $\theta_1 = \theta_e^{\mathrm{new}}$. If these conditions are not met, we repeat this procedure until the desired tolerance is achieved. 

    We investigate the mapping~\eqref{eq:mapping} along with the previously described algorithm to identify successive collisions with the boundary for $\gamma = 3$, varying the values of $\xi$. We examine 150 randomly selected initial conditions, iterating each one for $N = 10^4$ times [Figure~\ref{fig:phase_space}]. The system exhibits a complex coexistence of regular and chaotic domains across all considered values of $\xi$, characteristic of quasi-integrable Hamiltonian systems. As $\xi$ increases, the chaotic domain expands, leading to the destruction of stability islands. The period-3 islands undergo multiple bifurcations and mutations. However, as $\xi$ approaches $0.85$ [Figure~\ref{fig:phase_space}(h)], several smaller islands emerge. Beyond this threshold, the system becomes ``less'' chaotic, \textit{i.e.}, the chaotic domain diminishes as $\xi \rightarrow 1$~\cite{billiard2}.

    \section{Survival probability}
    \label{sec:survivalprobability}

\begin{figure}[tb]
        \centering
        \includegraphics[width=\linewidth]{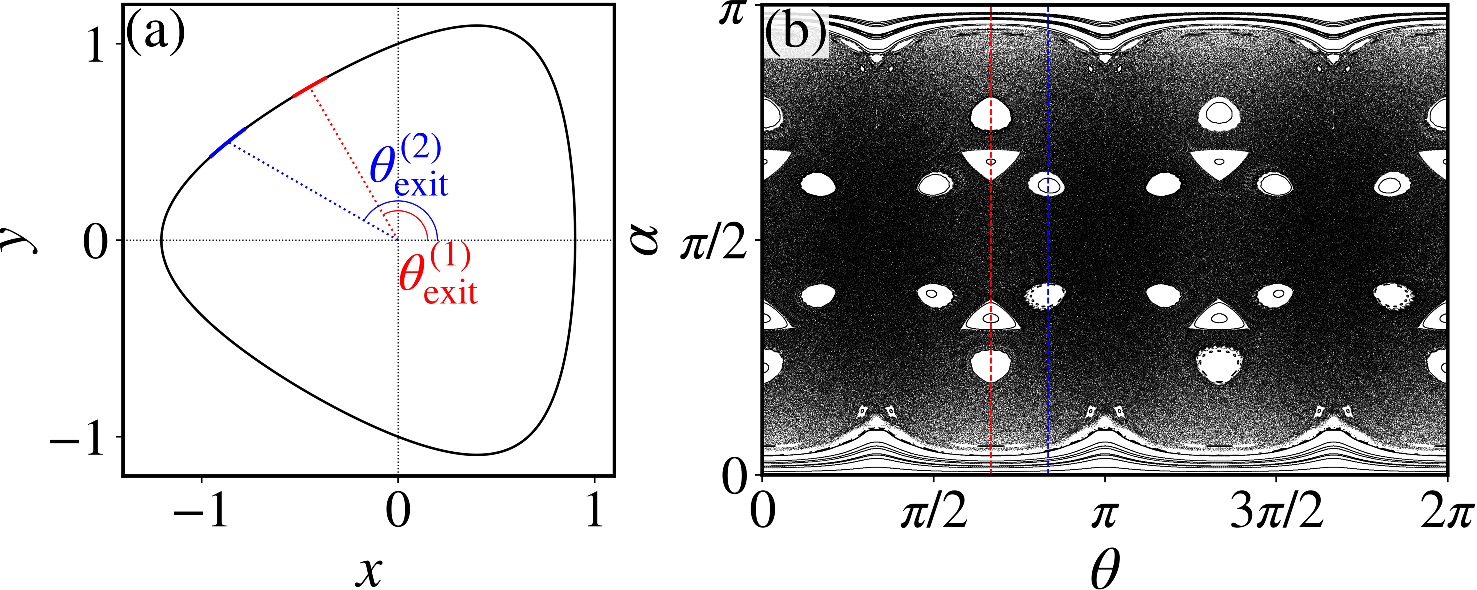}
        \caption{(a) The billiard boundary and (b) the phase space for $\gamma = 3$ and $\xi = 0.45$. The red ($\theta_{\mathrm{exit}}^{(1)} = 2\pi/3$) and blue ($\theta_{\mathrm{exit}}^{(2)} = 5\pi/6$) lines in (a) on the boundary represent the holes with size $h = 0.20$. The dashed lines in (b) correspond to the positions in phase space where the holes are centered.}
        \label{fig:exits}
    \end{figure}

    In this section, we explore the properties for the escape of particles through a hole of size $h$, measured in the polar angle units, positioned on the billiard boundary. We consider $\gamma = 3$ and $\xi = 0.45$. Initially, we choose two distinct hole locations (Figure~\ref{fig:exits}) centered at $\theta_{\mathrm{exit}}^{(1)} = 2\pi/3$ and $\theta_{\mathrm{exit}}^{(2)} = 5\pi/6$. We initialize an ensemble of $M = 10^6$ randomly chosen particles within the phase space region defined by $(\theta, \alpha) \in [0, \pi/3]\times[\pi/2 - 0.25, \pi/2 + 0.25]$ and iterate each particle up to $N = 10^6$ collisions \footnote{For the chosen parameter values, there are only a few very small stability islands within this region (not shown). If a particle happens to fall within one of these small islands, we exclude it from the analysis. This exclusion does not significantly impact the overall behavior of the survival probability, as the number of discarded particles is much smaller than the total number of particles in the ensemble.}. We keep only one hole open at a time and every time a particle collides with the exit, it escapes and we interrupt the simulation and initialize another particle. We repeat this procedure until the whole ensemble is exhausted. We compute the survival probability, $P(n)$, that corresponds to the fraction of particles that have not yet escaped through the hole until the $n$th collision. Mathematically, it is defined as
    \begin{equation}
        P(n) = \frac{1}{M}N_{\mathrm{surv}}(n),
    \end{equation}
    where $M$ is the total number of particles and $N_{\mathrm{surv}}(n)$ is the number of particles that have survived until the $n$th collision. It is widely known that for strongly chaotic systems, the survival probability decays exponentially~\cite{LEONEL20121669,Livorati2014,Mendez-Bermudez_2015,LIVORATI2018225} as
    \begin{equation}
        \label{eq:pn}
        P(n) \sim \exp\qty(-\kappa n)
    \end{equation}
    where $\kappa > 0$ is the escape rate. However, the stickiness effect affects the statistical properties of the escape of particles. For systems with mixed phase space, the decay is slower. It has been shown that for such systems, the decay is either a power law~\cite{Altmann2009,Livorati2012,BORIN2023113965} or a stretched exponential~\cite{DETTMANN2012403,deFaria2016,LIVORATI2018225}. Due to the stickiness effect, particles might be trapped near stability islands and resonance zones for a long, but finite, time leading to long escape times and causing the aforementioned deviations from the exponential decay.

    \begin{figure}[tb]
        \centering
        \includegraphics[width=\linewidth]{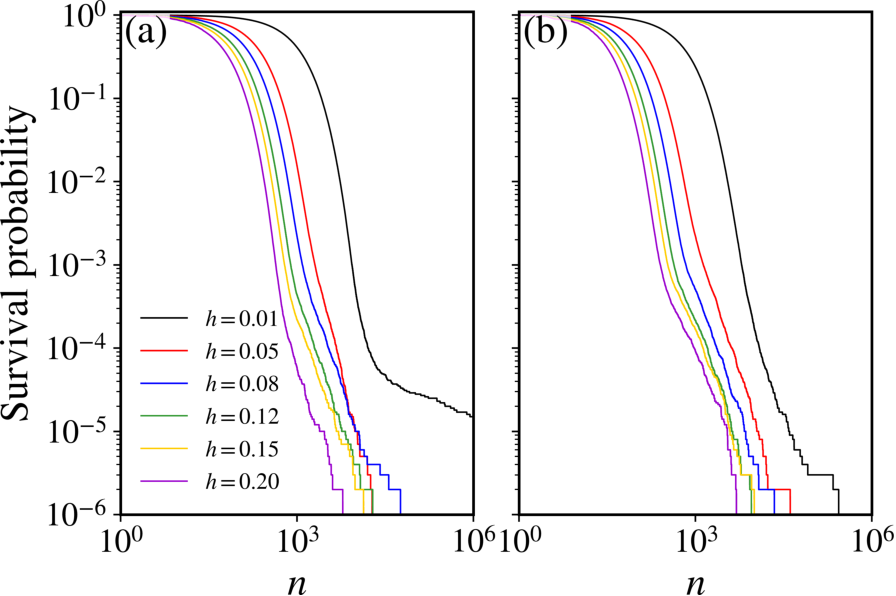}
        \caption{The survival probability through holes (a) \#1 ($\theta_{\mathrm{exit}}^{(1)} = 2\pi/3$) and (b) \#2 ($\theta_{\mathrm{exit}}^{(2)} = 5\pi/6$) individually, for $\gamma = 3$, $\xi = 0.45$ and different values of $h$. We considered an ensemble of $M = 10^6$ initial conditions randomly distributed in $(\theta, \alpha) \in [0, \pi/3]\times[\pi/2 - 0.25, \pi/2 + 0.25]$.}
        \label{fig:survprob_ALL}
    \end{figure}
    \begin{figure}[tb]
        \centering
        \includegraphics[width=\linewidth]{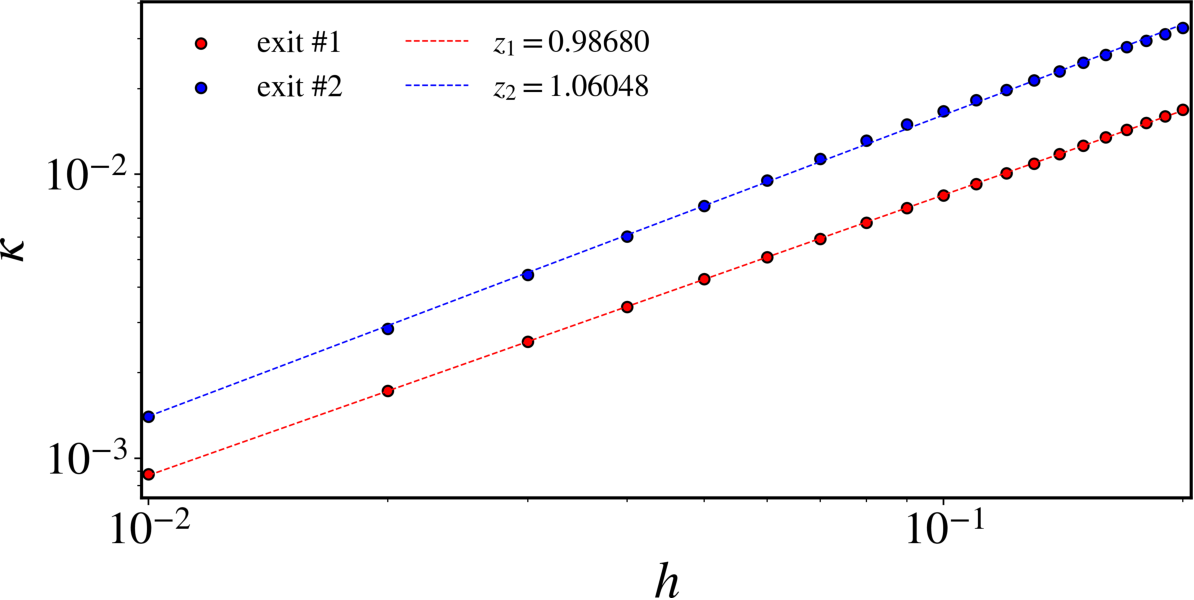}
        \caption{(a) The escape rate for holes (red dots) \#1 and (blue dots) \#2 as a function of the hole size $h$. The dashed lines correspond to the optimal fit based on the function $f(h) \sim h^z$.}
        \label{fig:escrate}
    \end{figure}

    \begin{figure}[tb]
        \centering
        \includegraphics[width=\linewidth]{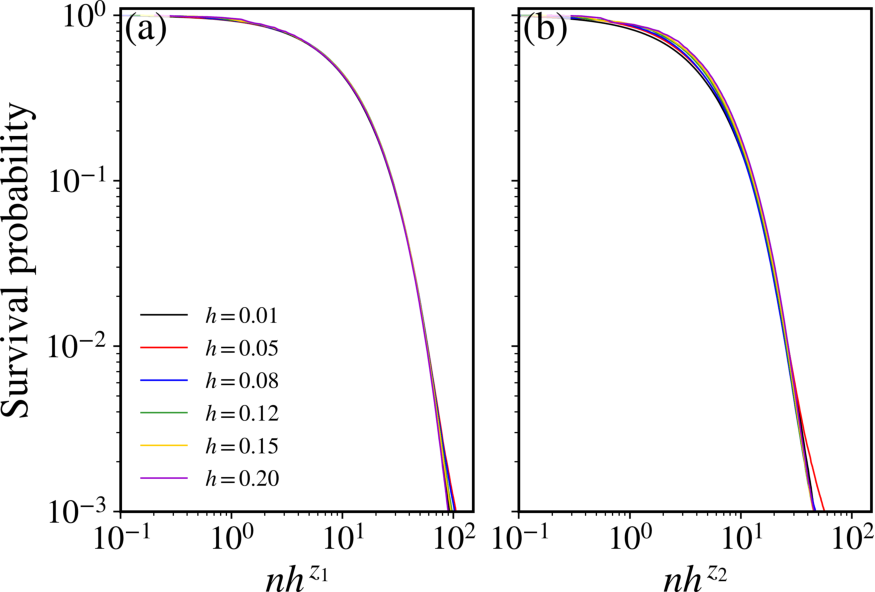}
        \caption{The survival probability through holes (a) \#1 and (b) \#2 individually, for $\gamma = 3$, $\xi = 0.45$ and different values of $h$ after the transformation $n \rightarrow nh^{z_i}$. Each $z_i$ corresponds to the value shown in Figure~\ref{fig:escrate}. The curves overlap onto a single and universal plot.}
        \label{fig:survprobSI_ALL}
    \end{figure}

    In Figure~\ref{fig:survprob_ALL}, we present the survival probability for six different hole sizes, $h$, calculated considering the two hole positions shown in Figure~\ref{fig:exits}. Both holes exhibit qualitatively similar behavior. For short times, the decay is exponential, while for longer times, a power-law tail emerges, which is a characteristic feature of the stickiness effect. Furthermore, $\kappa$ depends on $h$ as a power law, $\kappa(h)\sim h^z$ (Figure~\ref{fig:escrate}), with exponents $z_1 = 0.98680$ and $z_2 = 1.06048$ for holes \#1 and \#2, respectively. The knowledge of these exponents allows us to rescale the horizontal axis by the transformation $n\rightarrow nh^{z_i}$ making the survival probabilities of the corresponding holes overlap onto a single, and hence, universal plot (Figure~\ref{fig:survprobSI_ALL}).

    \begin{figure}[t]
        \centering
        \includegraphics[width=\linewidth]{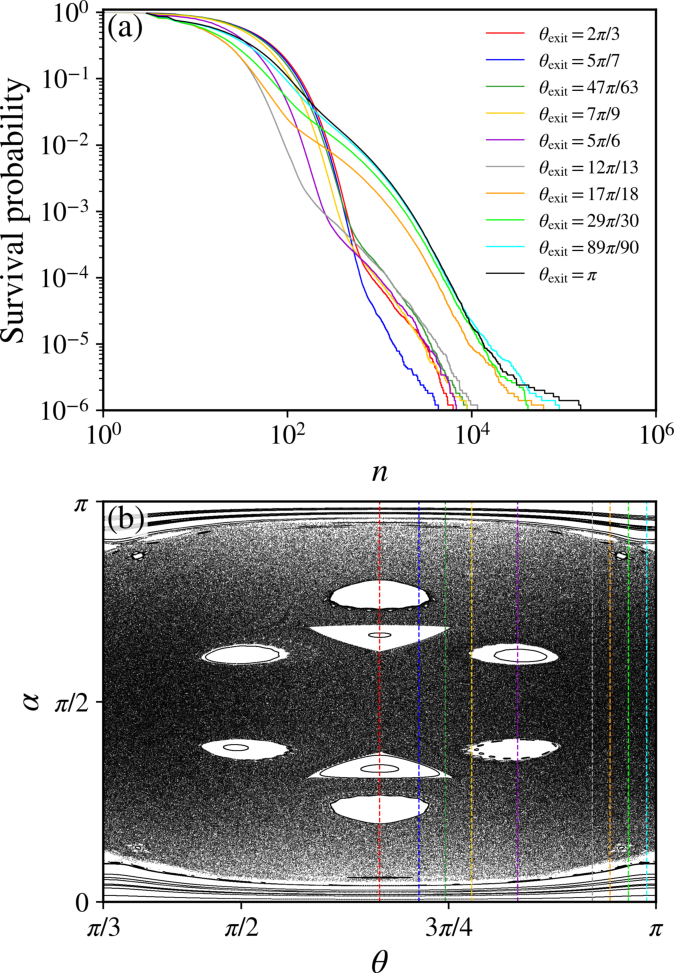}
        \caption{(a) The survival probability for $\gamma = 3$, $\xi = 0.45$ and $h = 0.20$ for different hole positions marked by colored dashed lines in (b). We considered an ensemble of $M = 10^6$ initial conditions randomly distributed in $(\theta, \alpha) \in [0, \pi/3]\times[\pi/2 - 0.25, \pi/2 + 0.25]$.}
        \label{fig:surv_prob_vs_pos}
    \end{figure}

    The escape rate is larger for larger $h$, as expected. This leads to the following question: Is there a preferential location to place the hole to enhance the escape of particles~\cite{Dettmann2011}? Insights have already been provided in Refs.~\cite{HANSEN20163634,HANSEN2018355} for a different billiard system, indicating that the escape is faster when the hole is placed in regions without stability islands. Here, we observe different behaviors in the survival probability decay depending on whether the hole overlaps with one of the larger stability islands. We consider $\gamma = 3$, $\xi = 0.45$ and $h = 0.20$ and change the hole position in the interval $\theta_{\mathrm{exit}} \in [\pi/3, \pi]$. Some holes are placed over regions with stability islands, while others are placed over regions dominated by the chaotic sea. We calculate the survival probability [Figure~\ref{fig:surv_prob_vs_pos}(a)] for each one of these hole positions [Figure~\ref{fig:surv_prob_vs_pos}(b)].
    
    When the hole is over regions with islands, we observe what we have previously reported: the decay is exponential for small times, whereas for larger times the power law emerges. On the other hand, when the hole is in the chaotic sea, \textit{i.e.}, away from the main islands, the decay is slowed down, and we observe stretched exponentials. The difference is mainly because when the hole is placed partially or entirely over an island, it might destroy all orbits in the vicinity of this island. In other words, it might destroy sticky regions and resonance zones that are responsible for slowing down the decay.

    \section{Escape basins}
    \label{sec:escapebasins}
    
    \begin{figure*}[tb]
        \centering
        \includegraphics[width=\linewidth]{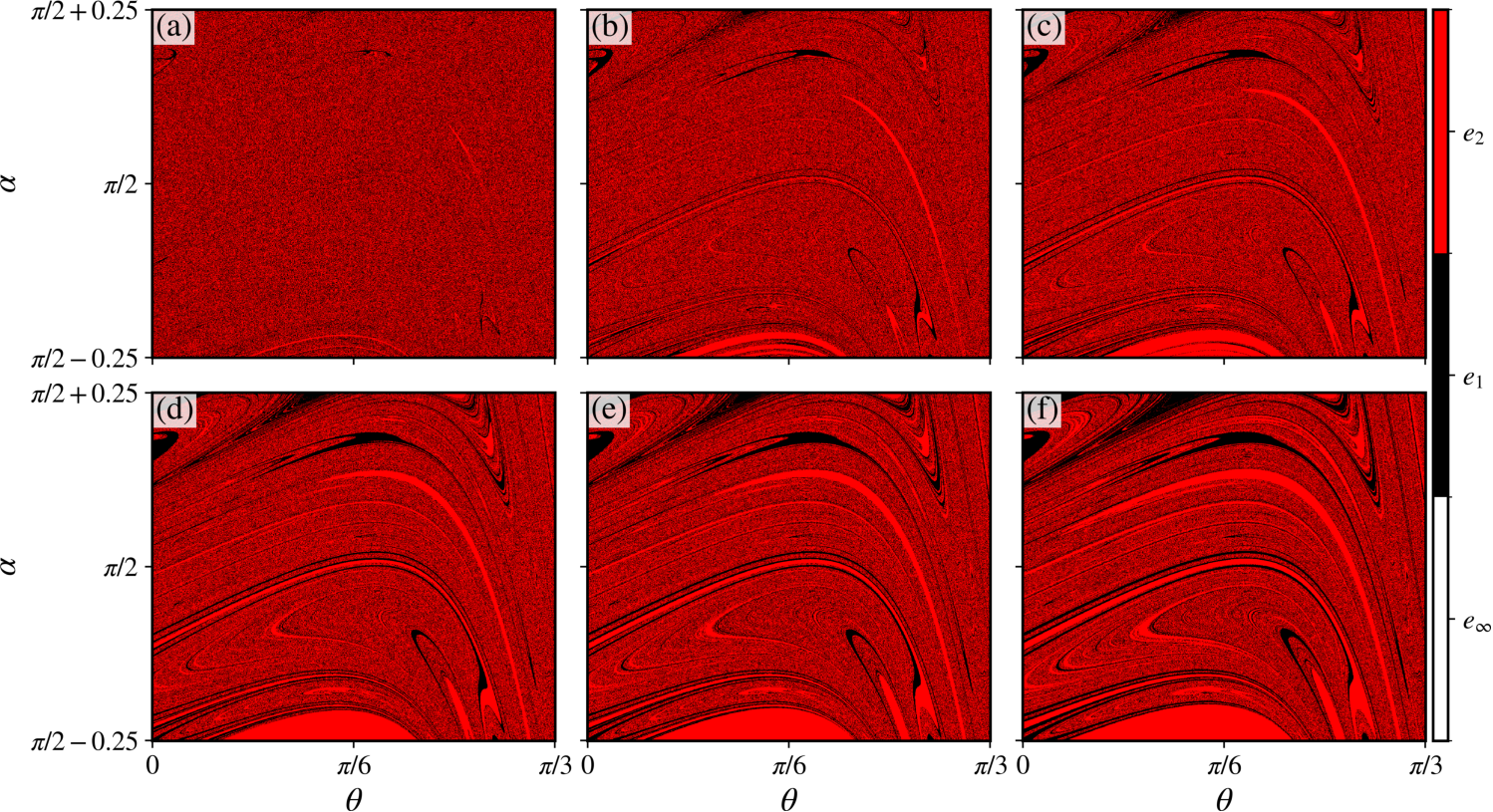}
        \caption{The escape basin for the particles that escape through holes (black) \#1 and (red) \#2 for $\gamma = 3$, $\xi = 0.45$ and (a) $h = 0.01$, (b) $h = 0.05$, (c) $h = 0.08$, (d) $h = 0.12$, (e) $h = 0.15$, and (f) $h = 0.2$.}
        \label{fig:escapebasin_2exits}
    \end{figure*}

    We have previously studied the escape of particles when one hole was open at a time. Next, we turn our attention to the escape dynamics when two holes are open simultaneously, and determine the escape basins for various hole sizes. We initialize an ensemble of $M = 10^6$ particles uniformly distributed in the phase space region delimited by $(\theta, \alpha) \in [0, \pi/3] \times [\pi/2 - 0.25, \pi + 0.25]$ for $\gamma = 3$ and $\xi = 0.45$. Each particle undergoes up to $N = 10^6$ collisions. We consider the same hole positions as in Section \ref{sec:survivalprobability} (Figure \ref{fig:exits}). To construct the escape basin, we iterate each particle until it escapes from one of the two exits. If a particle escapes from hole \#1 (\#2), we color the corresponding point black (red). If a particle does not escape within the maximum number of iterations, we color the point white. Figure~\ref{fig:escapebasin_2exits} shows the escape basins for six different hole size values $h$ when two holes are open. 
    
    For small hole sizes [Figure~\ref{fig:escapebasin_2exits}(a)], the black and red points are distributed almost at random, with nearly no discernible structure in the basin. As the hole sizes increase [Figure~\ref{fig:escapebasin_2exits}(b)-\ref{fig:escapebasin_2exits}(f)], the basins begin to exhibit a highly complex structure, characteristic of fractal basins. In order to quantify this complex structure, we apply the concept of basin entropy introduced by Daza and coworkers~\cite{Daza2016,Daza2017}. This method has been successfully applied to a variety of problems in nonlinear dynamics, such as dissipative~\cite{Mugnaine2021} and area-preserving~\cite{Mugnaine2018,Mathias2019,Souza2023} nontwist systems, drift motion of charged plasma particles~\cite{MATHIAS2017681,Souza2023b}, chaotic scattering in Hamiltonian sytems~\cite{Zotos2017,Nieto2020} as well as relativistic scattering~\cite{Bernal2018}. The basin entropy has been used to determine the fractal dimension of boundaries as well~\cite{PUY2021105588, GUSSO2021111532, Daza_2023}.

    The basin entropy quantifies the degree of uncertainty of a basin due to the fractality of the basin boundary. In order to calculate it, let us consider a bounded phase space region $\mathcal{R}$ which contains $N_A$ distinguishable asymptotic states. We discretize $\mathcal{R}$ into a mesh of $N_T \times N_T$ boxes of linear size $\delta$, and define an application $C:\mathcal{R}\rightarrow\mathbb{N}$ relating each initial condition to its asymptotic state. Daza \textit{et al.}~\cite{Daza2016} called this application a color. Each box contains a large number $N_p$ of initial conditions, each one leading to one of the $N_A$ colors (asymptotic states). For each box $i$, we associate a probability $p_{ij}$ of a color $j$ to exist in this box and define the Shannon entropy of the $i$th box as
    \begin{equation}
        \label{eq:shannon}
        S_i = - \sum_{j = 1}^{n_i}p_{ij}\log_2{p_{ij}},
    \end{equation}
    where $n_i \in [1, N_A]$ is the number of different colors inside the $i$th box. The probability $p_{ij}$ is simply the ratio between the number of points with color $j$ and the total number of colors (initial conditions) in the box. In this paper, we consider a box with 25 initial conditions and cover the phase space region with $216 \times 216$ boxes, totalizing $1080^2 = 1166400$ initial conditions.

    \begin{figure}[tb]
        \centering
        \includegraphics[width=\linewidth]{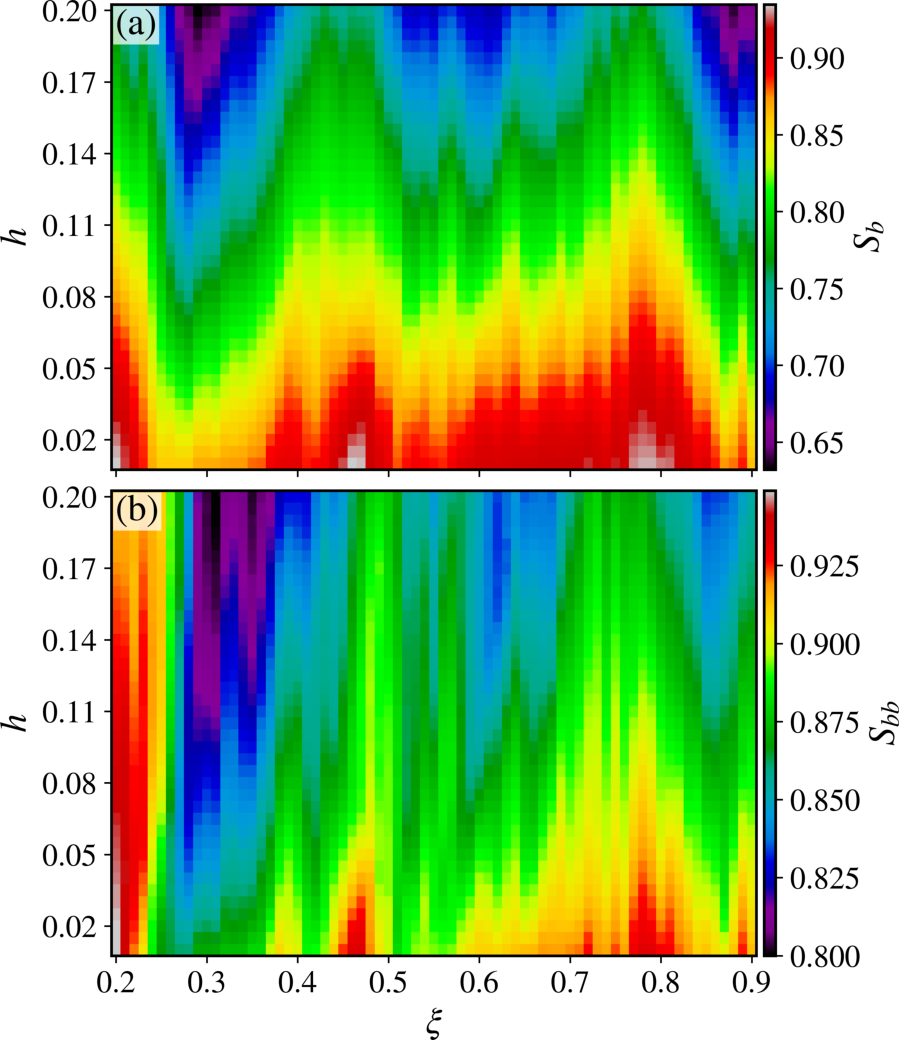}
        \caption{The basin entropy, $S_b$, and the basin boundary entropy, $S_{bb}$, for the escape basin considering two holes (Figure~\ref{fig:escapebasin_2exits}) as a function of $\xi$ and the hole sizes $h$ with $\gamma = 3$.}
        \label{fig:sbsbb2}
    \end{figure}

    \begin{figure*}[t]
        \centering
        \includegraphics[width=\linewidth]{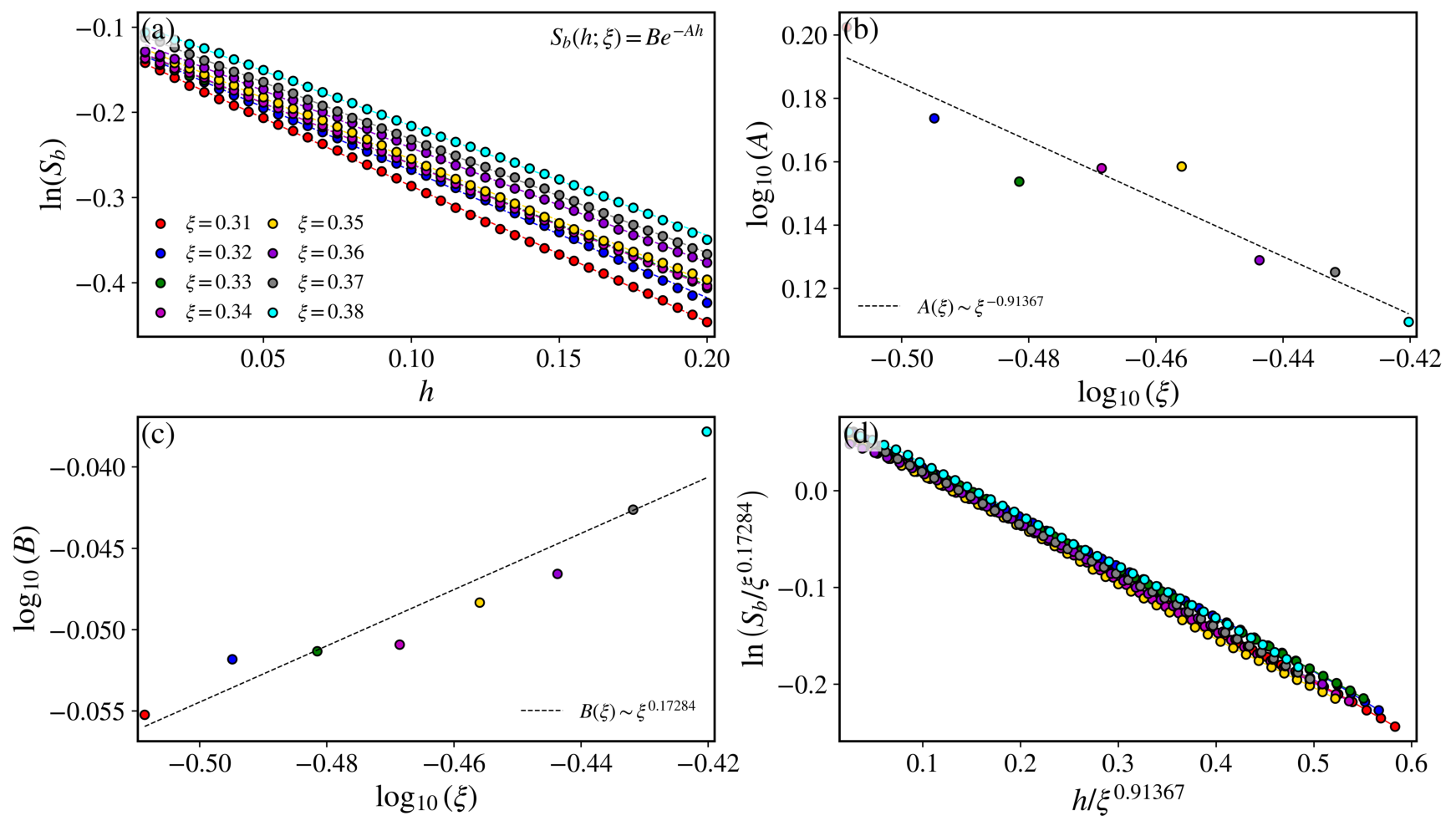}
        \caption{(a) The basin entropy, $S_b$, as a function of the holes size, $h$, for different values of $\xi$ (colored dots). The dashed lines correspond to the optimal fit based on the function $S_b(h; \xi) = Be^{-Ah}$. (b) and (c) The coefficients $A$ and $B$ obtained from the fitting in (a) as a function of $\xi$. Both coefficients scale with $\xi$ as a power law and the dashed lines correspond to the optimal fit based on the function $f(\xi) \sim \xi^{\zeta}$. (d) The overlap of $S_b$ onto a single and universal plot after the transformations $h\rightarrow h/\xi^{0.91367}$ and $S_b \rightarrow S_b/\xi^{0.17284}$.}
        \label{fig:SI2exits}
    \end{figure*}

    If the boxes covering $\mathcal{R}$ are nonoverlapping, the entropy of the phase space region is simply the sum of the entropies of all boxes
    \begin{equation}
        S = \sum_{i = 1}^{N_T^2}S_i,
    \end{equation}
    and the basin entropy $S_b$ and the basin boundary entropy $S_{bb}$ are defined as
    \begin{equation}
        \begin{aligned}
            S_b &= \frac{S}{N_T^2},\\
            S_{bb} &= \frac{S}{N_b},
        \end{aligned}
    \end{equation}
    where $N_b$ is the number of boxes that contain more than one color. The basin entropy, $S_b$, measures the basin degree of uncertainty, \textit{i.e.}, for a single asymptotic state $S_b = 0$, whereas for $N_A$ equiprobable asymptotic states, $S_b$ has its maximum value of $S_b = \log_2N_A$. On the other hand, the basin boundary entropy, $S_{bb}$, measures the uncertainty related only to the basin boundary. A fractality criterion has been provided by Daza \textit{et al.}~\cite{Daza2016}: if $S_{bb} > \log_22 = 1$, then the boundary is fractal. However, this is a sufficient but not necessary condition. In other words, if $S_{bb} > 1$, the boundary is fractal, however, if the boundary is fractal, $S_{bb}$ might not satisfy this condition.

    In our case, there are only three possible asymptotic states: the particles escape from either hole \#1 or hole \#2 or it does not escape at all (up to $10^6$ collisions), hence $N_A = 3$. To calculate the entropies, we determine the escape basins for $\gamma = 3$ and $\xi \in [0.2, 0.9]$ for different hole sizes in the interval $h \in [0.01, 0.20]$ (Figure~\ref{fig:sbsbb2} and Video 1 from the Supplementary Material). The basin entropy for small $h$ is large, as expected since the basins exhibit almost no ordered structure, as shown in Figure~\ref{fig:escapebasin_2exits}(a). As $h$ increases, structures start to appear, and $S_b$ and $S_{bb}$ decreases, in general, but in a non-trivial fashion. This leads to an important question about such a measure: \textit{Does the behavior of the basin entropy $S_b$ remain invariant under variations in $\xi$ and $h$?} To address our inquiry, we first plot the basin entropy $S_b$ as a function of the hole size $h$ for different $\xi$ values [Figure~\ref{fig:SI2exits}(a)]. We notice that $S_b$ is described by an exponential function of the form $S_b(h;\xi)=Be^{-Ah}$. The analysis of the coefficients $A$ and $B$ as a function of $\xi$ [Figures~\ref{fig:SI2exits}(b) and~\ref{fig:SI2exits}(c)] reveals a power law scaling for both of them, \textit{i.e.}, $A(\xi) \sim \xi^{\zeta_1}$ and $B(\xi) \sim \xi^{\zeta_2}$, with $\zeta_1 = -0.91367$ and $\zeta_2 = 0.17284$. Armed with these exponents, we rescale the horizontal and vertical axis by the transformations $h\rightarrow h/\xi^{-\zeta_1}$ and $S_b \rightarrow S_b/\xi^{\zeta_2}$, respectively. These transformations align the curves in Figure~\ref{fig:SI2exits}(a) onto a single, and hence, universal plot [Figure~\ref{fig:SI2exits}(d)], indicating that $S_b$ maintains its behavior regardless of $\xi$ within the chosen interval.

    \section{Final remarks}
    \label{sec:finalremarks}

    We have examined the statistical properties of the escape of particles from a billiard system by introducing a hole on the billiard boundary. Firstly our analysis focused on the behavior of the survival probability, which gives us information regarding the fraction of particles that have not yet escaped from the billiard up to a certain number of collisions. We have demonstrated that when the hole overlaps with the larger stability islands, the survival probability obeys an exponential decay, whereas when the hole is placed in a region dominated by the chaotic see, the decay follows a stretched exponential. Furthermore, in the cases where the hole is placed partially or entirely over a stability island, the survival probability exhibits scaling invariance with respect to the size of the exit, \textit{i.e.}, the survival probability preserves its behavior regardless of the hole size.

    Secondly, we have constructed escape basins for several values of the control parameter $\xi$ and the hole sizes $h$ by introducing two holes simultaneously. We have demonstrated that for $h \ll 1$, the basins exhibit an almost random pattern, with very few definite structures. As $h$ increases, the basins become increasingly complex with the emergence of multiple structures. In order to measure the complexity of these structures, we have applied the concept of basin entropy. We have found that both the basin entropy, $S_b$, and the basin boundary entropy, $S_{bb}$, are larger for smaller $h$ for all values of $\xi$, as expected. Moreover, they decrease as $h$ increases. The relation between these entropies and the parameters $\xi$ and $h$ is highly irregular and non-trivial. However, we have found that the basin entropy does maintain its behavior under parameter variations for a specific parameter interval. For $\xi \in [0.31, 0.38]$, the basin boundary exhibits an exponential decay with $h$, $S_b(h;\xi) = Be^{-Ah}$, and the coefficients $A$ and $B$ scale with $\xi$ as a power law, with exponents $\zeta_1 = -0.91367$ and $\zeta_2 = 0.17284$. Upon rescaling the horizontal and vertical axis by $h \rightarrow h/\xi^{-\zeta_1}$ and $S_b \rightarrow S_b/\xi^{\zeta_2}$, respectively, we have demonstrated that the basin entropy curves align into a single, and universal, curve. This indicates that $S_b$ is robust under variations of $\xi$. We would like to emphasize that while our basin entropy analysis mainly focused on boxes with 25 initial conditions, we also conducted simulations with 9, 16, 36, and 64 initial conditions. These simulations produced similar results, showing only minor variations in the exponents $\zeta_1$ and $\zeta_2$. Due to this, we have chosen not to display them on this paper.
    
    As a perspective of future works, we intend to study this billiard system with time-dependent holes as well as with a time-dependent boundary.

    \section*{Declaration of competing interest}
    
    The authors declare that they have no known competing financial interests or personal relationships that could have appeared to influence the work reported in this paper.

    \section*{Code availability}

    The source code to reproduce the results reported in this paper is freely available on the Zenodo archive \cite{zenodo} and in the GitHub repository \cite{github}.

    \section*{Acknowledgments}

    This work was supported by the Araucária Foundation, the Coordination of Superior Level Staff Improvement (CAPES), the National Council for Scientiﬁc and Technological Development (CNPq), under Grant Nos. 01318/2019-0, 309670/2023-3, and by the São Paulo Research Foundation, under Grant Nos. 2019/14038-6, 2022/03612-6, 2023/08698-9.

%


\end{document}